\documentclass[]{aiaa-tc}% insert '[draft]' option to show overfull boxes

\usepackage{varioref}%  smart page, figure, table, and equation referencing
\usepackage{dcolumn}%   decimal-aligned tabular math columns
\newcolumntype{d}{D{.}{.}{-1}}
\usepackage{nomencl}%   nomenclature generation via makeindex
\makenomenclature
\usepackage{makeidx}
\makeindex
\usepackage{fancyvrb}%  extended verbatim environments
\fvset{fontsize=\footnotesize,xleftmargin=2em}
\usepackage{lettrine}%  dropped capital letter at beginning of paragraph
\usepackage{amsmath,amsfonts,amsthm,amssymb,mathtools}
\usepackage[colorlinks]{hyperref}
\hypersetup{pdfborder={0 0 0},
	colorlinks=true,
	linkcolor=black,
	citecolor=blue,
	urlcolor=blue
	}
\usepackage{tikz}
	
 \title{Influence of Waviness on the Elastic Properties of Aligned Carbon Nanotube Polymer Matrix Nanocomposites}

 \author{
  Itai Y. Stein%
    \thanks{Graduate Student, Department of Mechanical Engineering, AIAA Student Member.}
  ,\
  and Brian L. Wardle%
    \thanks{Professor, Department of Aeronautics and Astronautics, AIAA Associate Fellow.}\\
  {\normalsize\itshape
   Massachusetts Institute of Technology, Cambridge, MA, 02139}\\
 }

 \AIAApapernumber{YEAR-NUMBER}
 \AIAAconference{Conference Name, Date, and Location}
 \AIAAcopyright{\AIAAcopyrightD{YEAR}}

\begin{document}

\maketitle

\begin{abstract}
  The promise of enhanced performance has motivated the study of one dimensional nanomaterials, especially aligned carbon nanotubes (A-CNTs), for the reinforcement of polymeric materials. While early work has shown that CNTs have remarkable theoretical properties, more recent work on aligned CNT polymer matrix nanocomposites (A-PNCs) have reported mechanical properties that are orders of magnitude lower than those predicted by rule of mixtures. This large difference primarily originates from the morphology of the CNTs that reinforce the A-PNCs, which have significant local curvature commonly referred to as waviness, but are commonly modeled using the oversimplified straight column geometry. Here we used a simulation framework capable of analyzing 10$^{5}$ wavy CNTs with realistic stochastic morphologies to study the influence of waviness on the compliance contribution of wavy A-CNTs to the effective elastic modulus of A-PNCs, and show that waviness is responsible for the orders of magnitude \textit{over-prediction} of the A-PNC effective modulus by existing theoretical frameworks that both neglect the shear deformation mechanism and do not properly account for the CNT morphpology. Additional work to quantify the morphology of A-PNCs in three dimensions and simulate their full elastic constitutive relations is planned.
\end{abstract}

\nomenclature{$a$}{Waviness amplitude, nm}%
\nomenclature{$A$}{Cross-sectional area of carbon nanotube, nm$^{2}$}%
\nomenclature{\textit{A-PNC}}{Aligned carbon nanotube polymer matrix nanocomposite}%
\nomenclature{\textit{A-CNT}}{Aligned carbon nanotube}%
\nomenclature{$CNT$}{Carbon nanotube}%
\nomenclature{$E$}{Elastic modulus, GPa}%
\nomenclature{$G$}{Carbon nanotube shear modulus, GPa}%
\nomenclature{$K$}{Effective spring constant, N/m}%
\nomenclature{$L$}{Separation of nodes that define a wavy carbon nanotube, nm}%
\nomenclature{$NF$}{Nanofiber}%
\nomenclature{$SEM$}{Scanning electron microscopy}%
\nomenclature{$V_{\mathrm{f}}$}{Carbon nanotube volume fraction, \%}%
\nomenclature{$Y$}{Carbon nanotube longitudinal modulus, TPa}%
\nomenclature{$w$}{Waviness ratio}%
\nomenclature[g$\lambda$]{$\lambda$}{Waviness wavelength, nm}%
\nomenclature[g$\xi$]{$\xi$}{Contribution of deformation modes}%

\printnomenclature[2cm]

\section{Introduction}

\lettrine[nindent=0pt]{T}{he} exciting electrical,\cite{Bezryadin2000,Mooij2006,Wang2010,Xu2008} thermal,\cite{Shen2010,Zhang2011,Balandin2011,Marconnet2013} and mechanical properties\cite{Wu2005,Chen2006,Wen2008} of nanowires, nanofibers, and nanotubes make them prime candidates for the design and manufacture of next-generation material architectures with tuned properties.\cite{DeVolder2013,Kauffman2008,Cao2009,Lu2007,Fan2009,Rogers2010,Liu2011} When organized into aligned nanofiber (NF) arrays, the fabrication of highly scalable nanostructured architectures with controlled properties becomes possible.\cite{DeVolder2013} To utilize the exceptional and highly anisotropic intrinsic properties of NFs in material solutions, many recent studied focused on the use of aligned NF arrays, especially aligned carbon nanotubes (A-CNTs), in nanocomposite structures, specifically aligned CNT polymer matrix nanocomposites (A-PNCs).\cite{Liu2011,Marconnet2011,Vaddiraju2009,Garcia2008,Wardle2008,Shin2010,Sepulveda2013,Ci2008,Handlin2013,Stein2015b,Carey2013} However, the properties reported by these previous works did not live up to the behavior predicted using current theoretical frameworks.\cite{DeVolder2013} Some of the main reasons why existing models cannot accurately predict the behavior of A-CNTs architectures, such as A-CNT arrays, are the various CNT morphology and proximity effects,\cite{Kauffman2008,Cao2009,Liu2011,Cebeci2014} which can strongly impact properties,\cite{Ginga2014,Stein2015a,Stein2015b} but are not well understood and cannot be adequately described in current-generation theoretical frameworks.

\begin{figure}[b!]
 \centering
 \includegraphics{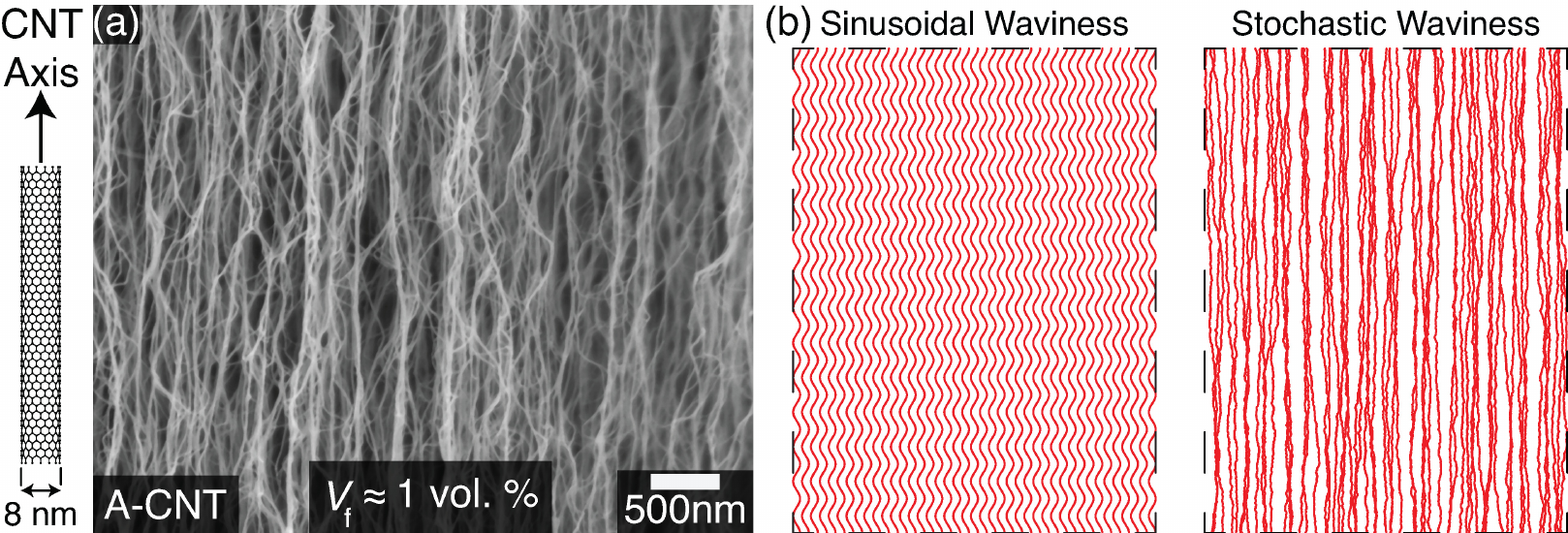}
 \caption{Waviness in carbon nanotube (CNT) arrays. (a) Scanning electron microscope (SEM) image of the cross-sectional morphology of a CNT array illustrating the random CNT waviness that results from the growth process. (b) Illustration of the sinusoidal (deterministic) waviness scheme, previously used in literature, and the stochastic (random) waviness scheme used here.}
 \label{fig1}
\end{figure}

Recent work has explored polymer nanocomposites comprised of both aligned (e.g. A-CNT arrays)\cite{Luo2007,Allen2013,Cebeci2009,CebeciPhD2011,Handlin2013,Herasati2014} and unaligned (e.g. dispersed CNT powders)\cite{Herasati2014,Shi2004,Shao2009,Shady2010,Rafiee2013,Dastgerdi2013,Dastgerdi2014,Varischetti2013,Fisher2002,FisherPhD2002,Fisher2003,Bradshaw2003,Anumandla2006,Pantano2010,Yazdchi2011,Matveeva2014,Montinaro2014,Paunikar2014,Li2008a,Omidi2010,Shokrieh2010,Yang2014} CNTs. The results of these previous studies, which used pure analytical (e.g. Mori-Tanaka model used in conjunction with the classic Eshelby solution),\cite{Luo2007,Allen2013,Shi2004,Shao2009,Shady2010,Rafiee2013,Dastgerdi2013,Dastgerdi2014} pure numerical (e.g. finite element models coupled with experiments),\cite{Varischetti2013,Herasati2014} or hybrids approaches of continuum micromechanics,\cite{Fisher2002,Fisher2003,Bradshaw2003,Anumandla2006,Pantano2010,Yazdchi2011,Matveeva2014,Montinaro2014,Paunikar2014} are in overall agreement that tortuous CNTs have significantly diminished reinforcement efficacy when compared to idealized straight (i.e. collimated) CNTs. Also, previous work on A-PNCs has shown that CNT waviness could lead to composite moduli that are $> 10\times$ lower than the ones predicted by rule of mixtures analysis of collimated CNTs,\cite{Cebeci2009,Handlin2013} and that the waviness of the CNTs can be reduced by CNT packing proximity, normally quantified by an increase in the CNT volume fraction ($V_{\mathrm{f}}$).\cite{Handlin2013,Stein2013} However, current-generation theoretical and numerical models cannot account for an evolving waviness ratio ($w$), and can only describe the CNT waviness using simple functional forms, e.g. sinusoidal or helical functional forms.\cite{Fisher2002,Fisher2003,Shady2010,Matveeva2014,Paunikar2014,Vainio2014} See Figure~\ref{fig1}b for an illustration of CNT waviness described using a sinusoidal scheme. These oversimplifications of the CNT morphology lead to large \textit{over-predictions} of the stiffness contribution of the CNTs to elastic modulus of the A-PNC, and a truly three-dimensional description of the CNT morphology that accounts the stochastic (random) nature of the CNT waviness (See Figure~\ref{fig1}b for illustration) and the evolution of $w$ with CNT packing proximity is necessary for more representative mechanical property prediction.\cite{Stein2015b}

In this work, a previously reported simulation framework capable of simulating 10$^{5}$ CNTs with stochastic waviness,\cite{Stein2015a,Stein2015b,Stein2015c} and representative scaling of $w$ with with $V_{\mathrm{f}}$, is used to evaluate the scaling of the intrinsic CNT elastic modulus ($E_{\mathrm{cnt}}$) with $w$.\cite{Stein2015a,Stein2015b} This is achieved by studying the contribution of (axial) stretching, (radial) shear, and bending on the deformation of CNTs in the A-PNC using an analysis similar to applied to study wavy CNTs,\cite{Stein2015a} which originates from early work on the mechanical behavior of carbon nanocoils.\cite{Chen2003} In conjunction with recent experimental data on an exemplary A-PNC system,\cite{Handlin2013} these results are used to compare the predicted scaling of the A-PNC elastic modulusp ($E_{\mathrm{pnc}}$) with $V_{\mathrm{f}}$ for the current scheme with the results of a previous finite element analysis (FEA) to show that more complete descriptions of the CNT morphology can lead to more accurate material property prediction. This report draws heavily from very recent studies on the mechanical behavior of wavy A-CNT (Ref.~\citenum{Stein2015a}) and wavy CNT polymer composites (Ref.~\citenum{Stein2015b}), to explore the mechanisms of CNT reinforcement in A-PNCs more in-depth.

\section{Methods}

Here we describe how the morphology and mechanical behavior of the CNTs, and subsequently A-PNCs, was quantified for an A-CNT system comprised of $\sim 8$ nm outer diameter multiwalled CNTs.

\begin{figure}[b!]
 \centering
 \includegraphics{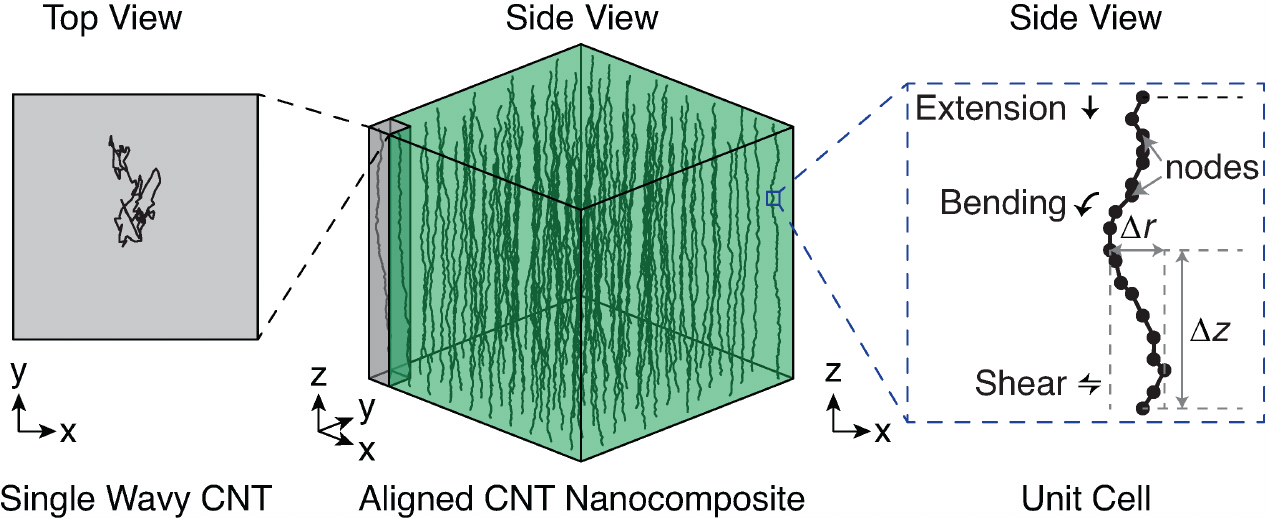}
 \caption{Top view of a simulated carbon nanotube (CNT) illustrating the two dimensional random walk that comprises the waviness (left), side view of an aligned carbon nanotube polymer matrix nanocomposite (A-PNC) comprised of simulated wavy CNTs (center), and illustration of the three CNT deformation modes that contribute to the elastic response of the wavy CNTs in the A-PNC (right).}
 \label{fig2}
\end{figure}

\subsection{Simulation Framework}

To simulate the morphology of CNTs in three dimensions, each CNT was discretized into an array of nodes in $xyz$ space. The width of the confining two dimensional area that bounds the node displacements was defined using the minimum and maximum inter-CNT spacings which correspond to the evolution of the CNT morphology with packing proximity.\cite{Stein2013,Stein2014b,Stein2015a} To apply the appropriate waviness to all other nodes, the displacement of each node relative to the node that precedes it, defined as $\Delta r$, was evaluated using the amplitude ($a$) extracted from the waviness ratio ($w$), and the node displacement increment in the $\hat{z}$ direction was set at a magnitude of $0.05 \lambda$, where $\lambda$ is the wavelength of the waviness ($\rightarrow \lambda = a/w$) that has a value equal to the maximum inter-CNT spacing,\cite{Stein2014b,Stein2015a,Stein2015b} so that a unit cell comprised of 10 nodes (see Fig.~\ref{fig2} for illustration) will have a total $\hat{z}$ displacement, defined as $\Delta z$, of magnitude $\lambda/2$. Stochastic waviness was achieved by using Gaussian distributions to independently evaluate the $x$ and $y$ displacements of the nodes using the $a$ extracted from $w$. $n \times n$ large A-CNT arrays were then assembled layer-by-layer, where each layer was comprised of $n$ wavy CNTs enclosed in their confining area, and the layers were arranged in a manner analogous to Bernal stacking ($i.e.$ ABAB type stacking) to ensure that the representative packing morphology, defined by the effective two dimensional coordination number,\cite{Stein2013} is satisfied. See Figure~\ref{fig2} for a top view illustrating the two dimensional random walk of a wavy CNT in square packing, where the confining area is a square, and the resulting simulated A-PNC comprised of wavy CNTs. Additional details can be found elsewhere.\cite{Stein2015a} To accurately predict the properties of the A-PNCs, the evolution of $w$ as a function of the $V_{\mathrm{f}}$ was explored.

\subsection{Waviness Evolution}

The waviness of the A-CNT arrays as a function of $V_{\mathrm{f}}$ ($\rightarrow w(V_{\mathrm{f}})$) was recently evaluated from SEM images of the cross-sectional morphology of A-CNT arrays using a simple sinusoidal amplitude-wavelength ($\rightarrow a/\lambda$) definition of the of $w$.\cite{Stein2015a} The recent study found that CNT confinement leads both the mean values and standard error of $w$ to decrease significantly from $\approx 0.20 \pm 0.02$ at $V_{\mathrm{f}} \approx 1\%$ CNTs to $\approx 0.10 \pm 0.01$ at $V_{\mathrm{f}} \approx 20\%$ CNTs, and the following scaling relation for $w(V_{\mathrm{f}})$ was reported:\cite{Stein2015a}

\begin{equation}
    w(V_{\mathrm{f}}) = \Lambda(a_{1}(V_{\mathrm{f}})^{b_{1}} + c_{1} \pm (a_{2}(V_{\mathrm{f}})^{b_{2}} + c_{2})/\sqrt{n}).
    \label{eqn1}
\end{equation}
where $a_{1} = -0.04967$, $b_{1} = 0.3646$, $c_{1} = 0.2489$ (coefficient of determination $\mathbb{R}^{2} = 0.9996$); $a_{2} = -0.0852$, $b_{2} = 0.2037$, $c_{2} = 0.2100$ ($\mathbb{R}^{2} = 0.9812$); $n = 30$ CNTs,\cite{Stein2015a} and $\Lambda$ is a scaling factor that represents a change in morphology ($\Lambda = 1$ signifies no change).\cite{Stein2015b} Since the magnitude and evolution of $w$ with $V_{\mathrm{f}}$ that is represented in Eq.~\ref{eqn1} is natively applicable to A-CNT arrays ($\rightarrow \Lambda = 1$), but may be altered by the infusion, and subsequent curing, of a polymer resin into the CNT array during the A-PNC manufacturing process, $\Lambda \neq 1$ here. This enables the evaluation of the impact of waviness on the mechanical properties of the A-PNCs, and the approximation of the influence of infusion polymer on the waviness of the CNTs, where recent work has shown that polymer infusion leads to reductions in CNT waviness in A-PNCs ($\rightarrow \Lambda < 1$).\cite{Stein2015b,Natarajan2015}

\subsection{Mechanical Modeling and Effective Nanocomposite Modulus}

The stiffness of wavy CNT is analyzed using the principle of virtual work similar to the recent analysis of the behavior of wavy CNTs arrays,\cite{Stein2015a} which was based on a previous study of the deformation of a carbon nanocoil.\cite{Chen2003} Since the polymer matrix will restrict the movement of the CNTs during deformation, the analysis used here assumes that deformation via torsion is minimal, consistent with the findings of a recent study of A-PNC deformation during compression~\cite{Carey2013}, and consists of three primary deformation mechanisms: extension, shear, and bending. See Figure~\ref{fig2} for an illustration of the three modes that contribute to the deformation of a wavy CNT in the A-PNC. This analysis is very similar to the one recently report on wavy CNTs,\cite{Stein2015a} where one unit cell, which is defined as a segment bound by two nodes in the $z$ direction and is evaluated numerically for each node of a wavy CNT, is used to quantify the contribution of the three deformation modes. Additional details can be found elsewhere.\cite{Stein2015a,Stein2015b} The extension ($\xi_{\mathrm{extension}}$), shear ($\xi_{\mathrm{shear}}$), and bending ($\xi_{\mathrm{bending}}$) contributions can be used with $V_{\mathrm{f}}$ to evaluate the effective spring constant ($K(w)$), intrinsic elastic modulus of the wavy CNTs ($E_{\mathrm{cnt}}(w)$), and the effective elastic modulus of an A-PNC ($E_{\mathrm{pnc}}(V_{\mathrm{f}})$) from rule of mixtures as follows:

\begin{figure}[b!]
 \centering
 \includegraphics{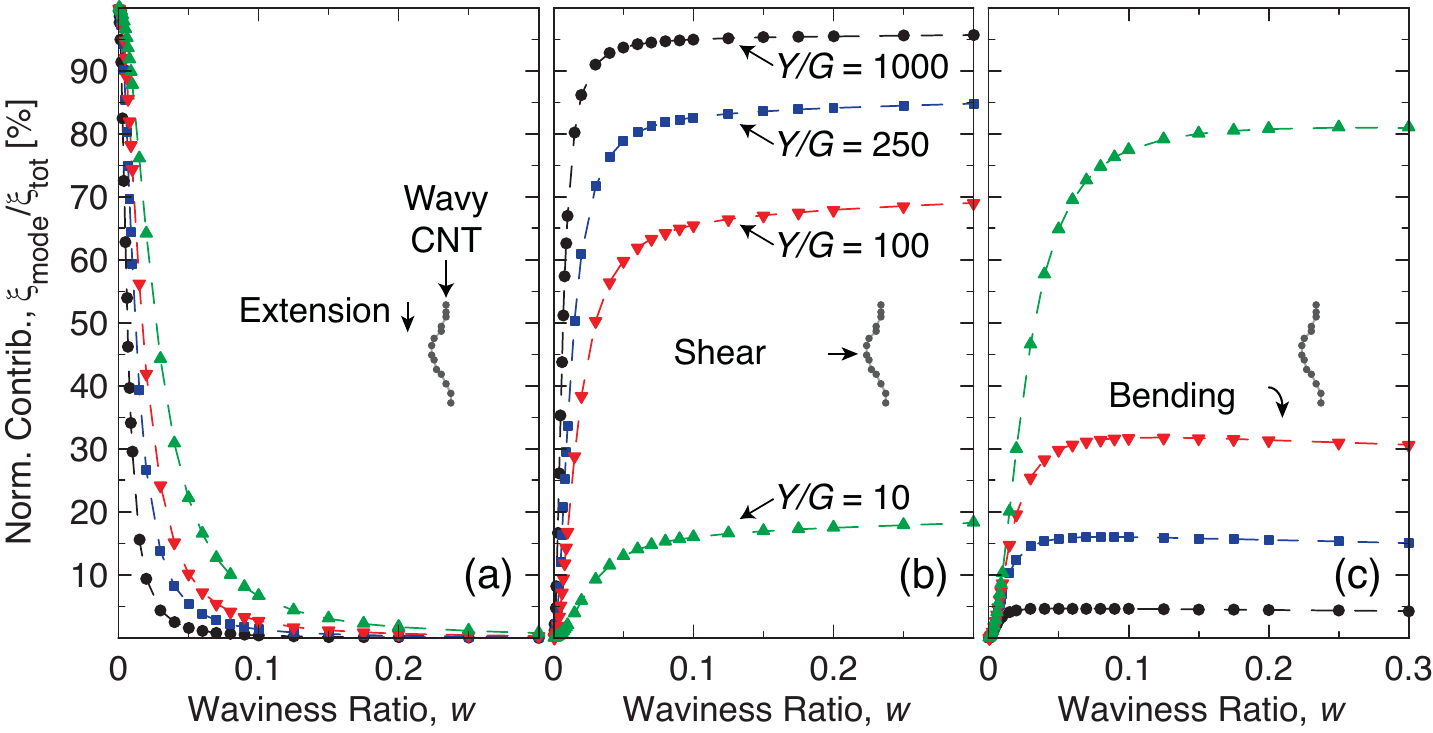}
 \caption{Contribution of extension, shear,and bending deformation mechanisms (see Eq.~\ref{eqn2a}) to the effective compliance ($\rightarrow \xi_{\mathrm{mode}}/\xi_{\mathrm{tot}}$ where $\xi_{\mathrm{tot}} = \sum\xi$) of wavy CNTs that comprise the A-PNC as a function of the waviness ratio ($w$) and ratio of the intrinsic CNT longitudinal and shear moduli ($\rightarrow Y/G$) for $Y/G = 1000$ (\protect\tikz{\protect\filldraw[draw=black,fill=black] (0,0) circle [radius=0.1cm];}), $Y/G = 250$ (\protect\tikz{\protect\filldraw[rounded corners=0.01cm,draw=blue,fill=blue] (0,0) rectangle (0.2cm,0.2cm);}), $Y/G = 100$ (\protect\tikz{\protect\filldraw[rounded corners=0.01cm,draw=red,fill=red,rotate=180](0,0)--(0.2cm,0)--(0.1cm,0.2cm)--cycle;}), and $Y/G = 10$ (\protect\tikz{\protect\filldraw[rounded corners=0.01cm,draw=green,fill=green](0,0)--(0.2cm,0)--(0.1cm,0.2cm)--cycle;}). (a) Extension mode contributions. (b) Shear mode contributions. (c) Bending mode contributions.}
 \label{fig3}
\end{figure}

\begin{subequations}
\label{eqn2}
\begin{equation}
    K(w) = (\xi_{\mathrm{extension}} + \xi_{\mathrm{shear}} + \xi_{\mathrm{bending}})^{-1}
    \label{eqn2a}
\end{equation}
\begin{equation}
    E_{\mathrm{cnt}}(w) = K(w)\left(\frac{L}{A}\right)
    \label{eqn2b}
\end{equation}
\begin{equation}
    E_{\mathrm{pnc}}(V_{\mathrm{f}}) = E_{\mathrm{cnt}}(w(V_{\mathrm{f}}))V_{\mathrm{f}} + E_{\mathrm{m}}(1 - V_{\mathrm{f}})
    \label{eqn2c}
\end{equation}
\end{subequations}
where $L$ is the separation of the two nodes in the $z$ direction (a measure that ensures that the average CNT tortuosity generated using $w$ in the stochastic waviness scheme is consistent with the CNT tortuosity that would from applying $w$ using a sinusoidal waviness scheme),\cite{Stein2015a} $A$ is the cross-sectional area of the CNTs (hollow cylinder geometry) evaluated using $D_{\mathrm{i}} \sim 5$ nm and $D_{\mathrm{o}} \sim 8$ nm,\cite{Lee2015,Stein2014a} and $E_{\mathrm{m}}$ is the elastic modulus of the polymer matrix in the A-PNCs studied here and in Ref.~\citenum{Handlin2013}. The analysis carried out here assumes that the longitudinal modulus ($Y$) modulus of the CNTs studied is constant and has a value of $Y \sim 1$ TPa,\cite{DeVolder2013,Peng2008} while the shear modulus ($G$) of the CNTs is governed by the elastic properties of the polymeric matrix, where $G \approx 1$ GPa when $G > E_{\mathrm{m}}$ (similar to pure A-CNT arrays),\cite{Stein2015a} and $G \approx E_{\mathrm{m}}$ when $G \leq E_{\mathrm{m}}$.\cite{Stein2015b} Since previous morphology characterization of the polymeric matrix in the A-PNCs found no evidence that CNT confinement leads to polymer morphology changes characteristic of the formation of an interphase region,\cite{Wardle2008,Cebeci2009} Eq.\ref{eqn2a}$-$\ref{eqn2c} assume that the polymer matrix and wavy CNTs are perfectly bound (i.e. perfect load transfer). Previous work in Ref.~\citenum{Needleman2010} showed that when the interphase size is very small and the CNT-polymer interfacial strength is $\gtrsim 150$ MPa, as reported by recent studies,\cite{Ozkan2012,Chen2013,Chen2015} the assumption of perfect bonding will have a very small impact on the predicted effective modulus of the A-PNCs. A recent study showed that for an interphase thickness of $\sim 1$ nm, the reinforcement modulus of the CNTs in the A-PNCs would be diminished by $\lesssim 3\%$ for $0.1 \leq w \leq 0.5$,\cite{Atescan2015} which also confirms that interphase effects will have a very weak influence on $E_{\mathrm{pnc}}(V_{\mathrm{f}})$ (from Eq.~\ref{eqn2c}).

\section{Results and Discussion}

\begin{figure}[t!]
 \centering
 \includegraphics{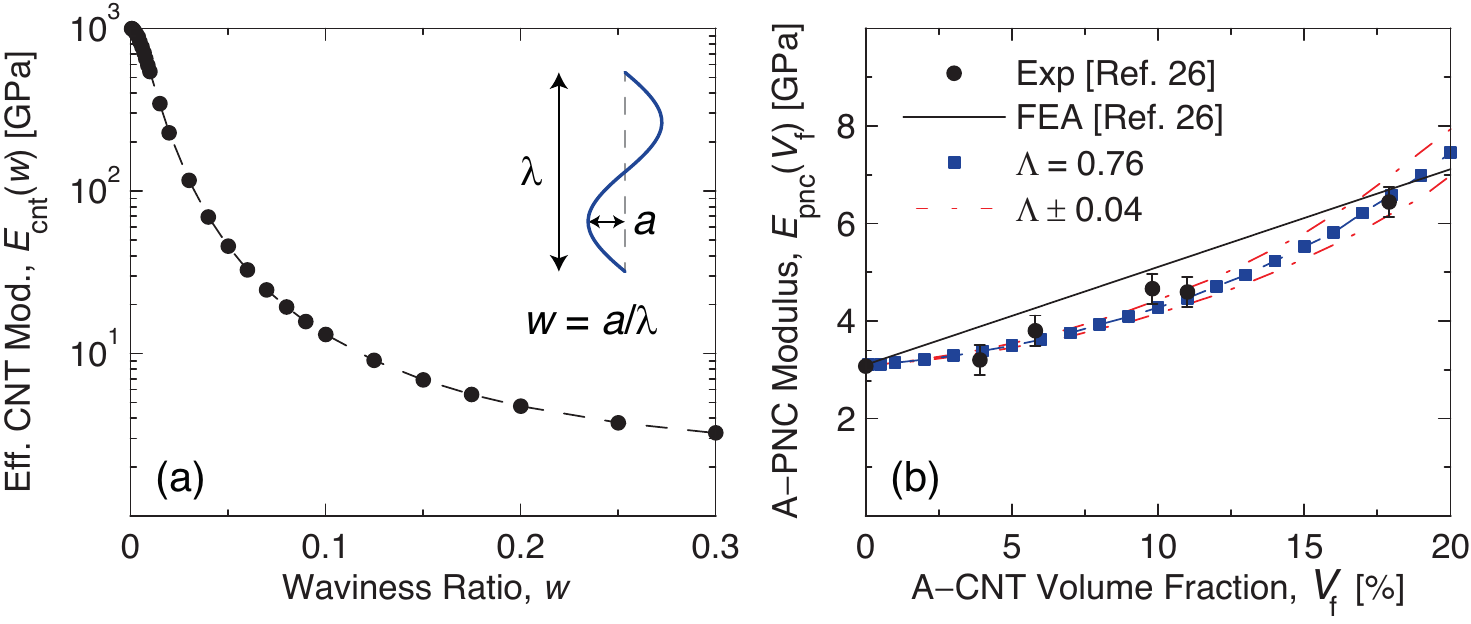}
 \caption{Mechanical properties of wavy carbon nanotubes (CNTs) and A-CNT polymer matrix nanocomposites (A-PNCs). (a) Plot of the intrinsic modulus of wavy CNTs ($E_{\mathrm{cnt}}(w)$) as a function of the waviness ratio ($w$) demonstrating that waviness can lead to orders of magnitude reductions in stiffness. (c) Plot comparing the scaling of the effective modulus of an A-PNC with the CNT volume fraction ($E_{\mathrm{pnc}}(V_{\mathrm{f}})$) for the 10$^{5}$ simulated wavy CNTs (via Eq.~\ref{eqn1} and Eq.~\ref{eqn2}) to the previously reported experimental (Exp) and finite element (FEA) scaling of $E_{\mathrm{pnc}}(V_{\mathrm{f}})$.\cite{Handlin2013} This plot shows that more representative descriptions of the CNT waviness can lead to more accurate mechanical property prediction for A-PNCs, and indicates that infusing A-CNTs with polymer to form the A-PNC causes a reduction in the CNT waviness.}
 \label{fig4}
\end{figure}

The elastic response of a CNT in the A-PNC is governed by the longitudinal, i.e. $Y$, and shear, i.e. $G$, effective CNT moduli. Since $G$ for the CNTs that comprise the A-PNC will likely not exceed the inter-plane modulus of graphite ($\sim30$ GPa), or be much smaller than $G \sim 1$ GPa previously reported for wavy CNTs,\cite{Stein2015a} $Y/G$ was varied from $Y/G = 10$ to $1000$ to represent the full of range of expected CNT effective stiffness contributions. See Figure~\ref{fig3} for the compliance contributions of each deformation mode ($\rightarrow \xi_{\mathrm{mode}}/\xi_{\mathrm{tot}}$ where $\xi_{\mathrm{tot}} = \sum\xi$). Figure~\ref{fig3}a illustrates that for $Y/G = 10$, 100, 250, and 1000, extension governs the elastic response of the CNTs at $w < 0.03$ by contributing $> 50\%$ of the effective CNT compliance. However, at $w > 0.03$, the anisotropy of $Y$ and $G$ determines the deformation mechanism that dominates the elastic response, where $Y/G = 10$ leads to bending contributing $> 50\%$ of the effective CNT compliance (Figure~\ref{fig3}c), while at $Y/G = 100$, 250, and 1000 the shear deformation mechanism contributes $> 50\%$ of the effective CNT compliance (Figure~\ref{fig3}b). Also, while bending is the most important mode for $Y/G < 100$, figure~\ref{fig3}b demonstrates that the shear contribution to the effective CNT compliance is $> 10\%$, meaning that previous analyses that only focused on bending and extension may yield predicted A-PNC moduli that are not representative.\cite{Fisher2002,Fisher2003} While recent simulation results are overall in agreement that the longitudinal stiffness of CNT arrays and their composites will be diminished by orders of magnitude as CNT waviness/curviness is increased,\cite{Herasati2014,Ginga2014,Dastgerdi2014,Matveeva2014,Montinaro2014,Paunikar2014,Yang2014}we show that the observed large stiffness losses originate from the large compliance contribution of the shear deformation mode governed by $G$ (see figure~\ref{fig3}b), and demonstrate that this effect can be mitigated by either choosing a stiffer matrix material ($\rightarrow$ smaller $Y/G$) or by decreasing the CNT waviness ($\rightarrow$ smaller $w$), the latter accomplished here by increasing the CNT $V_{\mathrm{f}}$.

As Figure~\ref{fig4}a demonstrates, CNT waviness significantly impacts their mechanical properties, and leads to a $100\times$ drop in $E_{\mathrm{cnt}}$ at $w \gtrsim 0.10$ ($\rightarrow E_{\mathrm{cnt}}(w = 0.10) \sim 10$ GPa), and $200\times$ drop in $E_{\mathrm{cnt}}$ at $w \gtrsim 0.20$ ($\rightarrow E_{\mathrm{cnt}}(w = 0.30) \sim 5$ GPa). This large change in $E_{\mathrm{cnt}}$ can be attributed to the very small value of $G$ of these CNTs ($\rightarrow G \sim 3.1$ GPa here), which is nearly three orders of magnitude smaller than $Y$ ($\rightarrow Y \sim 1$ TPa, so $Y/G \approx 323$ here),\cite{DeVolder2013,Peng2008,Stein2015b} and causes the shear deformation mechanism to dominate the mechanical behavior of the wavy CNTs that reinforce the A-PNC (see Figure~\ref{fig3}b). Using $w(V_{\mathrm{f}})$ (see Eq.~\ref{eqn1}), Eq.~\ref{eqn2} was used to predict $E_{\mathrm{pnc}}$ with $V_{\mathrm{f}}$ ($\rightarrow E_{\mathrm{pnc}}(V_{\mathrm{f}})$). See Figure~\ref{fig4}b for a plot of $E(V_{\mathrm{f}})$ as a function of $w(V_{\mathrm{f}})$ using 10$^{5}$ simulated wavy CNTs ($\rightarrow$ standard error of $\lesssim 0.5\%$). As Figure~\ref{fig4}b demonstrates, the scaling of $w$ has a very strong impact on $E(V_{\mathrm{f}})$, where $\Lambda \sim 0.76 \pm 0.04$ agrees very well with the previously reported experimental values of $E(V_{\mathrm{f}})$.\cite{Handlin2013} This indicates that polymer infusion leads to $\sim 25\%$ reductions in $w$ in A-PNCs, which is consistent with the $\sim 10\% - 25\%$ reductions in $w$ reported in recent studies.\cite{Stein2015b,Natarajan2015} In the previous study, the observed enhancement in stiffness was explained through a finite element analysis (FEA) for wavy CNTs using a sinusoidal description of the waviness (See Figure~\ref{fig4}b).\cite{Handlin2013} The FEA results illustrated that the waviness of the CNTs in the A-PNCs appears to decrease significantly as $V_{\mathrm{f}}$ is increased, but since the previous simulation could only analyze CNTs with a constant $w$,\cite{Handlin2013} the scaling of $E_{\mathrm{pnc}}$ with $V_{\mathrm{f}}$ could not be accurately reproduced. The ability of the simulation scheme to use representative descriptions of the magnitude and evolution of the CNT waviness when modeling $E_{\mathrm{cnt}}$ can lead to more accurate predictions of the mechanical behavior of A-PNCs (via $E_{\mathrm{pnc}}$) as a function of the CNT morphology (via $w$) and packing (via $V_{\mathrm{f}}$), which could enable precise tuning and optimization of the mechanical properties of A-PNCs for a variety of applications.

\section{Conclusion}

In conclusion, a simulation framework that enables a representative description of the magnitude and evolution of the carbon nanotube (CNT) waviness was used to predict the mechanical behavior of an aligned CNT (A-CNT) polymer matrix nanocomposite (A-PNC). The simulation results indicate that the CNT waviness, quantified via the waviness ratio ($w$), is responsible for more than $100\times$ reduction in the effective CNT stiffness. Also, by including information on the evolution of $w$ with the CNT volume fraction, the simulation is able to replicate the experimentally measured (A-PNC) modulus,\cite{Handlin2013} and outperform the mechanical property predictions of a previous finite element analysis that was only capable of analyzing CNTs with a constant $w$. Additionally, the simulation indicates that the presence of a polymer matrix reduces the waviness of the A-CNTs in the A-PNC by $\approx 24 \pm 4\%$ when compared to the waviness of as-grown (un-reinforced) A-CNT arrays. Further work to elucidate the origin of the observed waviness reduction is require, and future study of the morphology of A-CNT arrays and A-PNCs via three-dimensional transmission electron microscopy is planned. Also, once additional information on the morphology of A-CNT arrays and A-PNCs in three dimensions in available, the full elastic constitutive relations of these architectures will be analyzed and simulated. Using this simulation framework, more accurate material property predictions for CNT and other nanofiber based architectures may become possible, potentially enabling the design and fabrication of next-generation multifunctional materials with controlled properties.

\section*{Acknowledgments}

This work was supported by Airbus Group, Boeing, Embraer, Lockheed Martin, Saab AB, TohoTenax, and ANSYS through MIT's Nano-Engineered Composite aerospace STructures (NECST) Consortium and was supported (in part) by the U.S. Army Research Office under contract W911NF-07-D-0004 and W911NF-13-D-0001, and supported (in part) by AFRL/RX contract FA8650-11-D-5800, Task Order 0003. I.Y.S. was supported by the Department of Defense (DoD) through the National Defense Science \& Engineering Graduate Fellowship (NDSEG) Program. The authors thank Diana Lewis (MIT), John Kane (MIT) and the entire necstlab at MIT for technical support and advice. This work made use of the core facilities at the Institute for Soldier Nanotechnologies at MIT, supported in part by the U.S. Army Research Office under contract W911NF-07-D-0004, and was carried out in part through the use of MIT's Microsystems Technology Laboratories.

\end{document}